\documentclass[aps,prl,twocolumn,a4paper,english]{revtex4}
\usepackage{amsmath,epsfig,amssymb,subfigure,bm,dsfont}
\usepackage{lipsum}

\begin{document}

\title{Supervised Learning by Chiral-Network-Based Photonic Quantum
Computing}
\author{Wei-Bin Yan$^{1}$, Ying-Jie Zhang$^{1}$\footnote{%
yingjiezhang@qfnu.edu.cn}, Zhong-Xiao Man$^{1}$\footnote{%
manzhongxiao@163.com}, Heng Fan$^{2}$\footnote{%
hfan@iphy.ac.cn}, and Yun-Jie Xia$^{1}$}
\affiliation{$^{1}$College of Physics and Engineering, Qufu Normal University, Qufu,
273165, China}
\affiliation{$^2$Beijing National Laboratory for Condensed Matter Physics, Institute of
Physics, Chinese Academy of Sciences, Beijing 100190, China}

\begin{abstract}
Benefiting from the excellent control of single photons realized by the
emitter-photon-chiral couplings, we propose a novel potential
photonic-quantum-computation scheme to perform the supervised learning
tasks. The gates for photonic quantum computation are realized by properly
designed atom-photon-chiral couplings. The quantum algorithm of supervised
learning, composed by integrating the realized gates, is implemented by the
tunable gate parameters. The learning ability is demonstrated by numerically
simulating the performance of regression and classification tasks.
\end{abstract}

\maketitle

\section{Introduction}

Machine learning aims to investigate the algorithms to learn from the data
and make predictions \cite{ML}. Benefiting from increasingly powerful
computer and algorithms, machine learning becomes a rapidly developing field
in computer science and has successful applications in many fields, such as
computer vision, pattern recognition, data mining, speech recognition,
natural language processing, and so on. Quantum computing realizes the
quantum algorithms based on the quantum mechanical phenomena, such as
quantum superposition and quantum entanglement. Thanks to the quantum
properties, quantum computing shows superiority in certain problems compared
to classical computers \cite{Nielsen,Shor,Grover}. It is expected to improve
machine learning by quantum computing \cite%
{QML,Biamonte,Hans,Rebentrost,QPCA,QGAN1,QBM,QFM,SLQ,blank,ker}, which is
known as quantum machine learning. Currently, it is a significant issue to
realize quantum machine learning based on the quantum hardware platform \cite%
{QCL}.

On the other hand, the platform known as chiral quantum optics \cite{chr}
has gained prominence in recent years. The chiral coupling between the
transversely tightly confined photons and the atoms with
polarization-dependent dipole transitions has been realized in practice \cite%
{Sollner,chr,Sayrin,Scheucher}. This is underpinned by the fact that the
transversely tightly confined photons, which propagate towards opposite
directions, gain different local polarizations. Chiral quantum optics
provides the exciting approaches to manipulate the photons. It is known that
single photons are one of the most suitable carriers for quantum information
because they transport at the speed of light and rarely interact with the
environment. Consequently, it will be of interest to develop the potential
photonic quantum computing platform based on chiral quantum optics. More
significantly, machine learning by quantum computing based on this potential
platform needs to be explored.

For these purposes, we develop photonic quantum computation based on the
platform of chiral quantum optics and then exploit a framework to perform
supervised learning tasks by the developed quantum computation. In photonic
quantum computation, it is critical to realize the efficient photonic gates
at the level of single quanta. Especially, it is difficult to realize the
two-qubit gate consisting of the operation on a single photon controlled by
another single photon due to the weak photon-photon interaction. In this
work, we represent a qubit by two different 1D waveguides. The two bases of
the qubit are denoted by the two spatial mode bases representing the single
photon transporting in different waveguides, respectively. The efficient
photonic rotation, phase, controlled-rotation and controlled-phase gates
operated at the single-photon level are realized by the emitter-waveguide
chiral couplings.%
\begin{figure}[t]
\includegraphics[width=8cm, height=5cm]{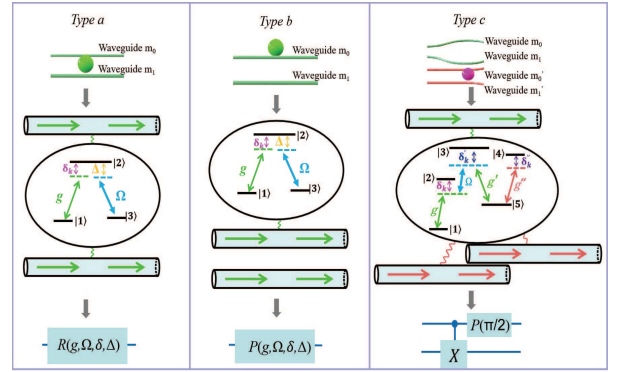}
\caption{Three types of waveguide-emitter chiral couplings constitute
single- and two-qubit photonic gates.}
\label{model}
\end{figure}
It is expected that the study opens an avenue for photonic
quantum computation based on the chiral quantum network \cite{cnw}.
Furthermore, we design a quantum algorithm to perform the supervised
learning tasks by properly integrating the realized gates into a circuit.
The performance of typical supervised learning tasks is numerically
simulated based on gradient descent optimization. Peculiarly, in our framework,
the matrix elements representing the single-photon gate show
nonlinear relationships with respect to the tunable gate parameters, which are
rewarded by the emitter-waveguide interaction. Thanks to this, the encoding
of the input data results in the complicated nonlinear mapping, which is known
as an essential role for the learning ability. Therefore, it is expected
that our work opens up an issue to develop the rich quantum feature mappings
\cite{QFM,QCL,ker} by the atoms with various structures interacting with
photons in quantum optics. The study represents a fresh potential quantum
optics platform for machine learning.

\section{Chiral-quantum-optics-based photonic quantum computation}

We first consider three types of atom-waveguide-chiral couplings as shown in
Fig. \ref{model}, which constitute essential gates for gate-based
photonic quantum computation. The calculations supporting the following main
outcomes are shown in the supplementary material.

\textbf{Photonic single-qubit rotation gate.} There is a $\Lambda $-type
atom chirally coupled to a pair of waveguides, \emph{type a} in Fig. \ref%
{model}. The atomic transition $\left\vert 1\right\rangle \leftrightarrow
\left\vert 2\right\rangle $ is coupled to the right-moving single photons in
the waveguide $m_{0}$ and waveguide $m_{1}$ with strength $g$. An external
Laser is introduced to drive the transition $\left\vert 2\right\rangle
\leftrightarrow \left\vert 3\right\rangle $. Initially, the atom is in its
ground state $\left\vert 1\right\rangle $. Due to the scattering, the single
photon moving in the waveguide $m_{0}$ ($m_{1}$) towards the atom is
delivered into the waveguide $m_{1}$ ($m_{0}$) with certain probability. The
transfer matrix corresponding to this process is
\begin{equation}
R(g,\Omega ,\delta _{k},\Delta )=\left[
\begin{array}{cc}
\frac{\delta _{k}\Delta _{k}-\Omega ^{2}}{\Delta _{k}(\delta _{k}-i\Gamma
)-\Omega ^{2}} & \frac{i\Delta _{k}\Gamma }{\Delta _{k}(\delta _{k}-i\Gamma
)-\Omega ^{2}} \\
\frac{i\Delta _{k}\Gamma }{\Delta _{k}(\delta _{k}-i\Gamma )-\Omega ^{2}} &
\frac{\delta _{k}\Delta _{k}-\Omega ^{2}}{\Delta _{k}(\delta _{k}-i\Gamma
)-\Omega ^{2}}%
\end{array}%
\right]  \label{tmat}
\end{equation}%
with $\Omega $ the Rabi frequency of the Laser, $\Gamma =g^{2}$, and $\Delta
_{k}=\delta _{k}-\Delta $. The symbol $\delta _{k}$ ($\Delta $) denotes the
detuning between the input photon (external Laser) and the corresponding
atomic transition.

The qubit is represented by the pair of waveguides. The facts that the
single photon with energy $E$ is in the waveguide $m_{0}$ and waveguide $%
m_{1}$ are described by the two degenerate orthogonal photonic quantum
states of the qubit, respectively.
The emitter-waveguide interaction plays
the role of a rotation on the two states and hence acts as a rotation gate.
In the quantum scattering theory \cite{Shen}, the photonic state is the free
state outside the emitter-waveguide interaction ranges. This represents the
advantage of the far less stringent operational requirements in the temporal
dimension, which can significantly reduce errors coming from the operation
in the temporal dimension.

\textbf{Photonic single-qubit phase gate.} There is a $\Lambda $-type atom
chirally coupled to the waveguide $m_{0}$, \emph{type b} in Fig. \ref{model}%
. It is the special case of \textit{type a} when the atom is decoupled to
the waveguide $m_{1}$. The single photon in the waveguide $m_{0}$ gains a
phase shift, with the phase tuned by the Laser, i. e.
\begin{equation}
P(g,\Omega ,\delta _{k},\Delta )=\left[
\begin{array}{cc}
\frac{\Delta _{k}(\delta _{k}+i\frac{\Gamma }{2})-\Omega ^{2}}{\Delta
_{k}(\delta _{k}-i\frac{\Gamma }{2})-\Omega ^{2}} & 0 \\
0 & 1%
\end{array}%
\right] .
\end{equation}

\textbf{Photonic two-qubit controlled-rotation and controlled-phase gates.}
There is a five-level atom chirally coupled to the waveguide $m_{1}$,
waveguide $m_{0}^{\prime }$ and waveguide $m_{1}^{\prime }$, \emph{type c}
in Fig. 1. The atomic transitions $\left\vert 1\right\rangle
\leftrightarrow \left\vert 2\right\rangle $ and $\left\vert 3\right\rangle
\leftrightarrow \left\vert 5\right\rangle $\ are driven by the right-moving
photons in the waveguide $m_{1}$ with strengths $g$ and $g^{\prime }$,
respectively. The transition $\left\vert 2\right\rangle \leftrightarrow
\left\vert 3\right\rangle $ is driven by an external Laser. The transition $%
\left\vert 4\right\rangle \leftrightarrow \left\vert 5\right\rangle $\ is
driven by the right-moving photon in the waveguide $m_{0}^{\prime }$ and
waveguide $m_{1}^{\prime }$ with strength $g^{\prime \prime }$. For
simplicity, we assume that all the atom-waveguide coupling strengths are
equal to $\sqrt{\Gamma }$ in our scheme. The atom is initialized in the
state $\left\vert 1\right\rangle $, and the $m$-th pair of waveguides
contains a right-moving photon.
We label the two situations that the photon
is contained in waveguide $m_{1}$ and in waveguide $m_{0}$ by \emph{situation A}
and \emph{situation B}, respectively.
In \emph{situation A}, the atom will absorb the
photon and then reemit it, meanwhile making the transition $\left\vert
1\right\rangle \rightarrow \left\vert 5\right\rangle $ with probability $%
P_{conv}$ or maintaining in the state\ $\left\vert 1\right\rangle $. The
former transition corresponds to the inelastic scattering in most cases
because of energy conservation. This constitutes a tunable frequency
convertor with the conversion efficiency $P_{conv}$. The frequency of
reemitted photon and the conversion efficiency can be tuned by the external
Laser. The quantum frequency convertor has many critical applications for
connecting the quantum systems with different frequencies. In this work, we
assume $\omega _{51}=\omega _{32}$, the input photon and the Laser
resonantly drive the atom, and the Rabi frequency of the Laser is equal to $%
\frac{\Gamma }{2}$. The symbol $\omega _{ij}$ denotes the transition
frequency between the levels $\left\vert i\right\rangle $ and $\left\vert
j\right\rangle $. In this case, the atom is determinably in the state $%
\left\vert 5\right\rangle $ after the elastic scattering. In addition, the
reemitted photon gains a phase shift of $\frac{\pi }{2}$. If another single
photon is subsequently injected into the waveguide $m_{0}^{\prime }$ ($%
m_{1}^{\prime }$), it drives the transition $\left\vert 4\right\rangle
\leftrightarrow \left\vert 5\right\rangle $.%
\begin{figure*}[t!]
\includegraphics[width=16cm, height=4.5cm]{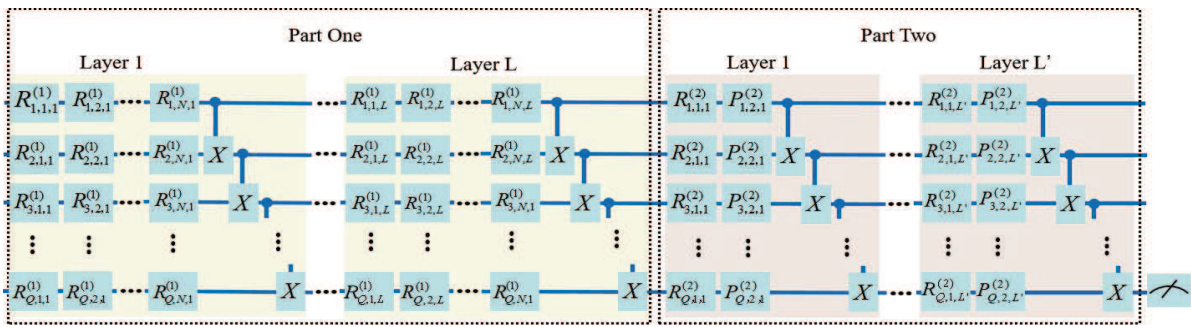}
\caption{The quantum circuit of machine learning. The gate $%
R^{(h)}_{j_{1},j_{2},j_{3}}$ ($P^{(h)}_{j_{1},j_{2},j_{3}}$)denotes the gate
represented by \textit{Type a} (\textit{Type b}), with the corner marks
representing different gates as shown in the main text. The gate $X$ is
composed by \textit{Type c}. Limited by the figure length, the $\frac{%
\protect\pi }{2}$-phase shift in \textit{Type c}, which is considered in the
numerical simulations, is not plotted.}
\label{learnin}
\end{figure*}
Then it is delivered into
waveguide $m_{1}^{\prime }$ ($m_{0}^{\prime }$) with the transfer matrix $%
\frac{\delta _{k}^{\prime \prime }}{\delta _{k}^{\prime \prime }-i\Gamma }I+%
\frac{i\Gamma }{\delta _{k}^{\prime \prime }-i\Gamma }X$. The operator $I$
denotes the two dimensional identity matrix, $X$ represents the Pauli-X gate,
and $\delta _{k}^{\prime \prime
} $ is the detuning between the single photon and the atomic transition $%
\left\vert 4\right\rangle \leftrightarrow \left\vert 5\right\rangle $. The
atom can transit to its initial state by interacting again with the
reemitted photon. A candidate for this is to bring in an auxiliary waveguide
connecting the waveguide $m_{1}$ with atoms of \textit{type a}. It is
possible that the reemitted photon is routed into the auxiliary waveguide
and then is delivered into the waveguide $m_{1}$ by switching the laser for
the long optical path.
In \emph{situation B}, the photon does not drive the atomic
transition and hence the atom maintains its initial state $\left\vert
1\right\rangle $. If another single photon is subsequently injected into the
waveguide $m_{0}^{\prime }$ ($m_{1}^{\prime }$), it is decoupled to the
atom. Therefore, \emph{type c} constitutes a combination of two operations.
One is the rotation operation on the photonic state of the $m^{\prime }$-th
pair of waveguides, which is controlled by the photonic state of the $m$-th
pair of waveguides. The other is the $\frac{\pi }{2}$-phase-shift operation
operated on the photon in the waveguide $m_{1}$. Especially, when the
transition $\left\vert 4\right\rangle \leftrightarrow \left\vert
5\right\rangle $ is decoupled to either of the $m^{\prime }$-th pair of
waveguides, the controlled-rotation gate becomes the controlled-phase gate.
In the following learning tasks, we consider the resonant case, i. e. $%
\delta _{k}^{\prime \prime }=0$, and hence the controlled-rotation gate
reduces to the C-NOT gate.

Therefore, the rotation, phase and controlled-rotation and controlled phase
gates can be realized by the chiral emitter-waveguide couplings. The
operation $R(g,\Omega ,\delta _{k},\Delta )$ can also be interpreted as the
single-qubit $X$ rotation, i. e. $R(g,\Omega ,\delta _{k},\Delta
)=e^{-i\frac{\theta }{2}}R_{X}(\theta )$, with $%
R_{X}(\theta )=\cos (\frac{\theta }{2})I-i\sin (\frac{\theta }{2})X $. The angle $%
\theta $, which satisfies $\sin \frac{\theta }{2}=\frac{-\Delta _{k}\Gamma }{%
\sqrt{(\Delta _{k}\delta _{k}-\Omega ^{2})^{2}+(\Delta _{k}\Gamma )^{2}}}$
and $\cos \frac{\theta }{2}=\frac{\Delta _{k}\delta _{k}-\Omega ^{2}}{\sqrt{%
(\Delta _{k}\delta _{k}-\Omega ^{2})^{2}+(\Delta _{k}\Gamma )^{2}}}$, can be
tuned from $0$ to $2\pi $ by the external Laser. Similarly, the operation $%
P(g,\Omega ,\delta _{k},\Delta )$ can be interpreted as the
single-qubit $Z$ rotation. More complicated gates can be obtained by the
integration of the realized gates. In practice, the imperfect chiral
couplings and the atomic dissipations to the other modes except for the
guided mode are harmful to the photonic gates. As shown in numerical
simulations in the supplemental material, the gates are efficient when the
Purcell factor $P\sim \frac{1}{60}$ \cite{puer} and $\frac{\Gamma _{R}}{%
\Gamma _{L}}\sim 50$ \cite{Sollner}, with $\Gamma _{L}$ denoting the
coupling strength of the emitter with the left-moving photon in the imperfect chiral
case.

\section{Chiral-quantum-optics-based supervised learning}

We proceed to develop an quantum algorithm to perform the supervised
learning tasks by integrating the realized photonic gates into a quantum
circuit, which can be considered as a chiral network. The quantum circuit,
as shown in Fig. \ref{learnin}, is composed by the realized gates operating
on $Q$ qubits.

The scheme is decomposed into two parts. Part One
contains $L$ layers. There are $N$ rotation gates, composed by \textit{type
a, }sequentially\textit{\ }operating on the $m$-th ($m=1...Q$) qubit in each
layer. Then the controlled-NOT gates composed by \textit{type c} are
performed to obtain the correlations between the nearest neighbor qubits.
Part Two contains $L^{\prime }$ layers. There are one rotation gate composed
by \textit{type a} and one phase gate\ composed by \textit{type b }acting on
the $m$-th qubit in each layer. Then the nearest neighbor qubits are
correlated by nonlocal operations, as done in Part One. To distinguish the
parameters belonging to different single-qubit gates of the global circuit,
we bring in the indexes to distinguish the different single-qubit gates. For example, we use
$\Omega _{j_{1},j_{2},j_{3}}^{(h)}$ to denote the Rabi frequency
corresponding to the $j_{2}$-th single-qubit gate, acting on the $j_{1}$-th
qubit, in layer $j_{3}$, part $h$.

For a given input set $\{\mathbf{x}_{i}\}$, we label the output function of
the learning model and the teacher data with $\{g(\mathbf{x}_{i},\mathbf{%
\Theta })\}$ and $\{f(\mathbf{x}_{i})\}$, respectively. In supervised
learning, one aims to update the parameter $\mathbf{\Theta }$ to make the
output function close to the teacher data. We consider that, initially, all
the $Q$ qubits are in the state $\left\vert 0\right\rangle $. After the
quantum operations represented in Fig. \ref{learnin}, the probability $P$
of the $Q$-th qubit in its state $\left\vert 1\right\rangle $ is
measured. Obviously, $P$ is the function with respect to the parameters of
the single-qubit gates. We assume that the atomic transition frequencies and
the frequencies of guided photons are fixed, while the frequencies and Rabi
frequencies of the external Lasers are adjustable. The adjustable parameters
are divided into two parts. One part is used to encode the input set and
hence labeled by $\mathbf{x}_{i}$. The other part contains the updated
parameters and is labeled by $\mathbf{\Theta }$. Then one can label $%
P\rightarrow P(\mathbf{x}_{i},\mathbf{\Theta ,F})$, with $\mathbf{F}$
denoting the set of the fixed gate parameters. In the following numerical
simulations, each of the detunings between the input photons and
corresponding atomic transitions is the rand number in the range $[-2\Gamma
,2\Gamma ]$ produced by the computer and then keeps constant. The output
function is defined by a linear transformation of $P$, i. e. $g(\mathbf{x}%
_{i},\mathbf{\Theta ,F})=\omega P(\mathbf{x}_{i},\mathbf{\Theta ,F})+b$,
with $\omega $ and $b$ real numbers. The quadratic cost function, $%
L=\frac{1}{2}\sum_{i}[g(\mathbf{x}_{i},\mathbf{\Theta ,F})-f(\mathbf{x}_{i})]^{2}$ is
introduced to measure the gap between the teacher data and the output
function. Based on the gradient descent, the parameters are iterated as $%
\mathbf{\Theta }\leftarrow\mathbf{\Theta }-\nabla _{\mathbf{\Theta }}L$
until the value of the cost function is small enough. The analytical
gradients of the measurement with respect to variational parameters can be
obtained based on the parameter-shift rules developed in Ref. \cite{Mari}
(see the supplementary material for details).

The input data $\mathbf{x}_{i}$ is encoded in the Laser
frequencies of the single-qubit rotation gates in Part One, i. e.
$\mathbf{x}_{i}=\mathbf{\Delta^{(1)}}$ with $\mathbf{\Delta^{(1)}}$ the set composed by all the detunings
$\Delta^{(1)}_{j_{1},j_{2},j_{3}}$ in Part One. The encoding
provides the complicated nonlinearity into the output function with respect
to the input data because the elements in the transfer matrix (\ref{tmat})
show the nonlinear relationship with $\Delta $. The tensor product structure
of the $Q$ qubits results in the tensor of the rotation matrices, which
brings in more complicated nonlinearity formally. The nonlocal operations
play a key role because the measurement of the $Q$-th qubit is independent
of the operations on other qubits if there is no nonlocal operation. These are
crucial for the learning ability. The system dimension increases
exponentially, i. e. $2^{Q}$, due to the tensor product structure. When the
dimension of the system is large enough, it is difficult to calculate by
classical computer. For the quantum system, the outcomes are obtained by
running the operations and performing the measurements. Quantum computing is
expected to perform efficient calculations when the dimension is large, which
is one purpose for most quantum algorithms.

\begin{figure}[t]
\centering\includegraphics[width=8cm, height=10cm]{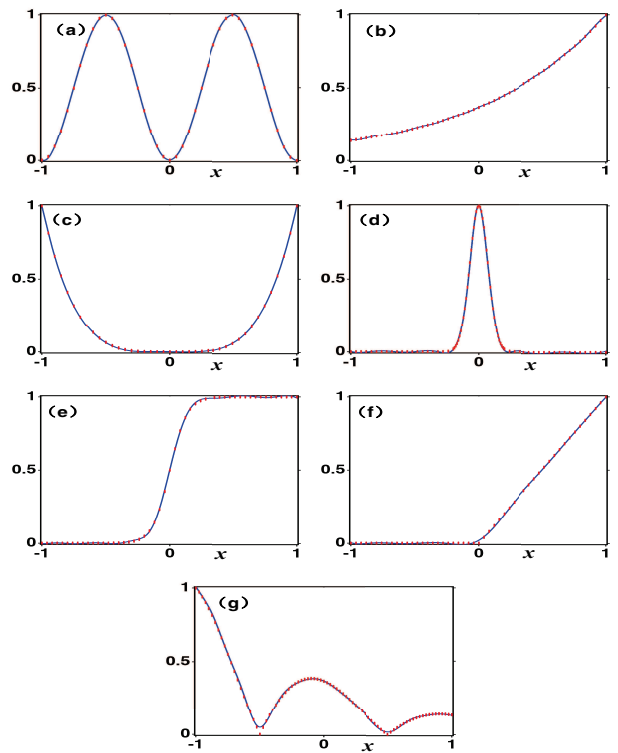}
\caption{Numerically simulating the performance of regression tasks. The red
dots denote the teachers $f(x)$ against the one-dimension input data $x$.
The blue solid lines denote the output functions of the learning model
against the encoding gate parameters, i. e. $\Delta^{(1)}_{a,b,c}$ with $%
a=1..Q $, $b=1..N$, and $c=1..L$. The parameters are $L=4$, $N=3$, $Q=4$, $%
L^{\prime }=8$, $\protect\omega =2$, $b=-0.5$, and $\Gamma$ is taken as a
unit. The target functions are: (a) $f(x)=sin^{2}(x\cdot \protect\pi )$, (b)
$f(x)=e^{x-1}$, (c) $f(x)=x^{4}$, (d) $f(x)=e^{-x^{2}/0.01}$, (e) $f(x)=%
\frac{1}{1+e^{-15x}}$, (f) Rectified Linear Unit, (g) $f(x)= |cos(x\cdot
\protect\pi )|e^{-(x+1)}$.}
\label{fig2}
\end{figure}

\begin{figure}[t]
\centering\includegraphics[width=8cm, height=8cm]{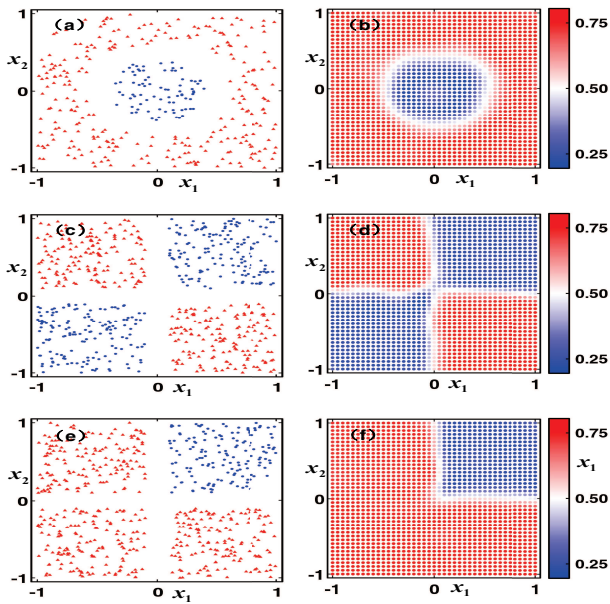}
\caption{Numerically simulating the performance of classification tasks.
(a), (c), and (e) represent the teachers. The blue dots indicate class $0$
with $f(x_{1},x_{2})=0$, while the red triangles indicate class $1$ with $%
f(x_{1},x_{2})=1$. The corresponding learning results, i. e. the output
function $g(x_{1},x_{2},\Theta)$ against the inputs, are represented in (b),
(d), and (f), respectively. The value of the output function that is smaller
(lager) than 0.5 indicates the class $0$ ($1$). The parameters are $L=4$, $%
N=6$, $Q=4$, $L^{\prime }=8$, $\protect\omega =1$, $b=0$, and $\Gamma$ is
taken as a unit.}
\label{fig3}
\end{figure}

From the recent viewpoint raised in \cite{ker}, the quantum operations of
Part One could be understood as a recent new concept of quantum feature
mapping \cite{QFM,QCL,ker}, and the measurement bases are altered by
variable parameters in Part Two. Obviously, if the $\Lambda $-type emitters
in Part One are replaced by the emitters with other structures,
different nonlinear mappings resulting from the encoding can be obtained. It
is possible to obtain rich quantum feature mappings by the atom with various
structures interacting with the photons. The interaction between the atom
with different structures and few photons is one of the research topics in
quantum optics. This implies the issue to develop the quantum feature
mapping by the platform of quantum optics.

We proceed to numerically simulate the performance of two prototypical
supervised learning tasks, i. e. regression and classification. The
performance of the regression tasks is shown in Fig. \ref{fig2}.
 For simplicity, we consider that all the gates are operated
in the ideal case. From the
encoding manner defined above, we take $\Delta _{a,b,c}^{(1)}=x$, with $%
a=1...Q$, $b=1...N$ and $c=1...L$. All the parameters belonging to $%
\mathbf{\Theta }$ are random numbers produced by the computer in the initial
time and then updated during the training in both the regression and
classification tasks. The variable parameters in Part Two adjust the rotation
angles of $R_{X}(.)$\ and $R_{Z}(.)$. The periodicity of the rotations
can reduce the values of the exploded gradients and meanwhile alters the gradient direction, which
efficiently reduce the possibility of the local minimum. In Fig. \ref{fig2}(a)-(d), the scheme well fits the typical
nonlinear functions. The target functions in Fig. \ref{fig2}(e) and (f) are
the typical activation functions used in neuron network. It implies that the
scheme may perform complicated learning tasks without bringing in the external
nonlinear function, such as the sigmoid function in the binary classification task. The target function in
Fig. \ref{fig2}(g) is a decay function, the similar lineshape to which is
common in quantum optics. For example, the evolution of the atomic
population in a structured environment with memory \cite{mem,mem2}. Machine
learning has been recently considered as a promising tool for the research
on quantum technology. It is expected that the scheme can serve quantum
technology. The simulations show that the framework can output a wide
variety of functions with respect to the simple input.

The numerical simulation of performing the binary classification tasks is
shown in Fig. \ref{fig3}. When the training input data is the $2$%
-dimensional vector, the $i$-th input data is denoted by $\mathbf{x}%
_{i}=(x_{i,1},x_{i,2})$. Then the gate parameters for encoding should be
divided into two halves. One encodes $x_{i,1}$ and the other encodes $%
x_{i,2} $. We assume that each layer in Part One contains $6$ single-qubit
rotation gates operating on the $m$-th ($m=1..Q$) qubit, i. e. $N=6$. The
input $x_{i,1}$ is encoded in the Laser frequencies of the 1st, 3rd and 5th
rotation gates while $x_{i,2}$ is encoded in the Laser frequencies of the
2nd, 4th and 6th rotation gates, i. e. $\Delta
_{j_{1},j_{2},j_{3},1}=x_{i,2-(j_{2}\ mod\ 2)}$ with $j_{2}\ mod\ 2$ the
remainder of $j_{2}\ $divided by $2$. The outcomes show that the framework
well performs different typical classification tasks.

\section{Discussions}

In summary, the photonic quantum computation has been realized based on
chiral quantum optics. Four elementary single- and two-qubit photonic gates
for photonic quantum computation are mediated by multi-level atoms chirally
coupled to the waveguides. More complex gates used in quantum computation
can be achieved by integrating the proposed gates. The tunable gate
parameters are composed by the parameters of the external Lasers, which
is employed to drive the emitters. Moreover, we develop the quantum machine
learning algorithm by integrating the realized gates into a quantum chiral
net. Some typical simple nonlinear supervised learning tasks are verified by
numerically simulating the performance with classical computer. During the
training, the gate parameters are tuned according to the gradient descent
optimization. It is reasonable that the scheme would be generalized to a
high-dimensional complex structure in practice, which is difficult for the
classical calculations, for various tasks. In our numerical simulations, the
ideal gates are considered for simplicity. In reality, the atomic decay and imperfect
chiral coupling influence the gate efficiency. Especially, the
latter, which brings difficulties to the simulation, would introduce
complicated interference\cite{yanwb}. Therefore, the imperfect chiral
coupling may improve the learning ability, which is hoped to be verified
experimentally in future. Currently, benefiting from the chiral
single-photon switch, a novel realization of the C-NOT gate operating on the
state of atom-like scheme is proposed \cite{zhouy}. The potential
realization of chiral photon emitter is analyzed in detail based on
different practical implementations, such as the optical photon interacting
with a NV center, the quantum dot interfaced to a semi-conductor waveguide,
and the guided mode composed by surface plasmon. It implies the potential
implementation for our scheme. Our work paves the way for the implementation
of photonic quantum computation, especially for quantum machine learning.

\begin{acknowledgments}
This work is supported by Taishan Scholar Project of Shandong Province
(China) under Grant No. tsqn201812059, the National Natural Science
Foundation of China (11505023, 61675115, 11647171, 11934018,12147146), and
the Strategic Priority Research Program of the Chinese Academy of Sciences
(XDB28000000).
\end{acknowledgments}

\end{document}


\title{Supplementary Material for ``Supervised Learning by
Chiral-Network-Based Photonic Quantum Computing''}
\author{Wei-Bin Yan$^{1}$, Ying-Jie Zhang$^{1}$, Zhong-Xiao Man$^{1}$, Heng
Fan$^{2}$, and Yun-Jie Xia$^{1}$}
\affiliation{$^{1}$College of Physics and Engineering, Qufu Normal University, Qufu,
273165, China}
\affiliation{$^2$Beijing National Laboratory for Condensed Matter Physics, Institute of
Physics, Chinese Academy of Sciences, Beijing 100190, China}
\maketitle

The supplementary material presents the detailed main calculations
supporting the conclusions in the main text.

\section{A $\Lambda $-type three-level atom is chirally coupled to two
waveguides}

\subsection{Transfer matrix}

\label{sec1} We consider a $\Lambda $-type three-level atom chirally coupled
to two waveguides at site $a$, \textit{type a} of Fig. 1 in the main text.
We assume that all the atom-waveguide strengths are frequency independent in
our scheme. This is equivalent to Weirdo-Winger approximation. When the
photonic frequency is away from the cut-off frequency, the dispersion is
approximately linearized. For conveniently describing the calculations, in
this section, we label the waveguide $m_{0}$ and waveguide $m_{1}$ by
waveguide $1$ and waveguide $2$, respectively. The frequencies corresponding
to the atomic states $\left\vert 1\right\rangle $, $\left\vert
2\right\rangle $, and $\left\vert 3\right\rangle $ are represented by $%
\omega _{1}$, $\omega _{2}$ and $\omega _{3}$, respectively. We choose the
ground-state energy for reference, i. e. $\omega _{1}=0$. The
time-independent Hamiltonian has the form

$$H =-i\sum_{j=1,2}\int dx_{j}r^{\dagger }(x_{j})\partial
_{x_{j}}r(x_{j})+\omega _{2}\sigma ^{22}+(\omega _{2}-\Delta )\sigma ^{33}
+\sum_{j=1,2}g\int dx_{j}\delta (x_{j}-a)\sigma ^{21}r(x_{j})+\Omega
\sigma ^{23}+h.c., \eqno(S1)$$
with $\sigma ^{mn}=\left\vert m\right\rangle \left\langle n\right\vert $ ($%
m,n=1,2,3$) the atomic raising, lowering and energy level population
operators. The operator $r^{\dagger }(x_{j})$($r(x_{j})$) creates(destroys)
a right-moving photon at site $x$ in waveguide $j$. The symbol $\Delta \ $%
denotes the detuning between the Laser and the corresponding atomic
transition. In this work, we take the photonic group velocity and Planck
constant unit. We assume that, initially, a right-moving single photon with
energy $E$ is injected into the waveguide from the left side and the atom is
in the state $\left\vert 1\right\rangle $. The arbitrary state of the system
is written as%
$$\left\vert \Psi \right\rangle =\int dx_{1}f_{1}(x_{1})r^{\dagger
}(x_{1})\left\vert \phi \right\rangle +\int dx_{2}f_{2}(x_{2})r^{\dagger
}(x_{2})\left\vert \phi \right\rangle +l\sigma ^{21}\left\vert \phi
\right\rangle +h\sigma ^{31}\left\vert \phi \right\rangle, \eqno(S2)$$
where $f_{1}(x_{1})$, $f_{2}(x_{2})$, $l$ and $h$ are corresponding
probability amplitudes. The state $\left\vert \phi \right\rangle $ denotes
that the atom is in the state $\left\vert 1\right\rangle $, and the two
waveguides do not contain any photon. The first and second terms denote the
states that there is a right-moving photon in waveguide $1$ and $2$,
respectively, meanwhile the atom is in its ground state $\left\vert
1\right\rangle $.

From the scattering eigenequation $H\left\vert \Psi \right\rangle
=E\left\vert \Psi \right\rangle $, we obtain%
$$ f_{1}(a^{+}) =f_{1}(a^{-})-i\sqrt{\Gamma }l,$$
$$f_{2}(a^{+}) =f_{2}(a^{-})-i\sqrt{\Gamma }l,$$
$$l =-\frac{\sqrt{\Gamma }}{2\delta _{k}}(f_{1}(a^{+})+f_{1}(a^{-}))-\frac{%
\sqrt{\Gamma }}{2\delta _{k}}(f_{2}(a^{+})+f_{2}(a^{-}))-\frac{\Omega }{%
\delta _{k}}h, $$
$$h =-\frac{\Omega }{\Delta _{k}}l,  \label{eigenset} \eqno(S3)$$

with $\delta _{k}=\omega _{2}-E$, $\Delta _{k}=\delta _{k}-\Delta $ and $%
\Gamma =g^{2}$. Then we derive the transfer matrix%
\begin{equation*}
\left[
\begin{array}{c}
f_{1}(a^{+}) \\
f_{2}(a^{+})%
\end{array}%
\right] =T\left[
\begin{array}{c}
f_{1}(a^{-}) \\
f_{2}(a^{-})%
\end{array}%
\right] ,
\end{equation*}%
with%
$$
T=\left[
\begin{array}{cc}
\frac{\Delta _{k}\delta _{k}-\Omega ^{2}}{\Delta _{k}(\delta _{k}-i\Gamma
)-\Omega ^{2}} & \frac{i\Delta _{k}\Gamma }{\Delta _{k}(\delta _{k}-i\Gamma
)-\Omega ^{2}} \\
\frac{i\Delta _{k}\Gamma }{\Delta _{k}(\delta _{k}-i\Gamma )-\Omega ^{2}} &
\frac{\Delta _{k}\delta _{k}-\Omega ^{2}}{\Delta _{k}(\delta _{k}-i\Gamma
)-\Omega ^{2}}%
\end{array}%
\right] \eqno(S4)
$$
In the main text, we use the label $R$ to denote the matrix because it acts
as a rotation.

When the atom is decoupled to the waveguide 2, the coupling is represented by%
\textit{\ type b }of Fig. 1 in the main text. The transfer matrix for this
case is obtained as%
$$
T^{\prime }=\left[
\begin{array}{cc}
\frac{\Delta _{k}(\delta _{k}+i\frac{\Gamma }{2})-\Omega ^{2}}{\Delta
_{k}(\delta _{k}-i\frac{\Gamma }{2})-\Omega ^{2}} & 0 \\
0 & 1%
\end{array}%
\right] .\eqno(S5)
$$%
The photon in waveguide 1 gains a phase shift due to the coupling. The phase
shift can be tuned by the Laser. It acts as a single-qubit phase gate.

\subsection{Analytical gradients of the measurement with respect to the
tunable gate parameters}

To perform the gradient descent optimization in the learning task, the
analytical gradients of the measurement with respect to tunable gate
parameters should be obtained. Here we represent the calculations of the
gradients to the tunable parameters in $T$ with the parameter-shift rules
developed in \cite{Mari}. The gradients to the tunable parameters in $%
T^{\prime }$ can be obtained in a similar way. Let us label an initial state
with density operator $\rho _{0}$. After the unitary transformation $%
KTK^{\prime }$, the output state is represented by $KTK^{\prime }\rho
_{0}K^{\prime \dagger }T^{\dagger }K^{\dagger }$, with the operators $K$ and
$K^{\prime }$ independent of the tunable parameters in $T$. We use $\rho
_{m} $ to denote the measurement bases. The measurement of the output is%
$$
m(\theta ) =Tr[KT(\theta )K^{\prime }\rho _{0}K^{\prime \dagger }T(\theta
)^{\dagger }K^{\dagger }\rho _{m}]
=Tr[(A+B\cos \theta +C\sin \theta ),\eqno(S6)$$

with $A=\frac{KK^{\prime }\rho _{0}K^{\prime \dagger }K^{\dagger }\rho
_{m}+KXK^{\prime }\rho _{0}K^{\prime \dagger }XK^{\dagger }\rho _{m}}{2}$, $%
B=\frac{KK^{\prime }\rho _{0}K^{\prime \dagger }K^{\dagger }\rho
_{m}-KXK^{\prime }\rho _{0}K^{\prime \dagger }XK^{\dagger }\rho _{m}}{2}$,
and $C=\frac{i(KXK^{\prime }\rho _{0}K^{\prime \dagger }K^{\dagger }\rho
_{m}-KK^{\prime }\rho _{0}K^{\prime \dagger }XK^{\dagger }\rho _{m})}{2}$.
The angle $\theta $ is represented in the main text. By the way, the
operation $T(\theta )$ satisfies $T(\theta \pm \phi )=T(\theta )T(\pm \phi )$%
. The gradient of the measurement with respect to $\theta $ is
$$
\partial _{\theta }m
=Tr[B\frac{\cos (\theta +s)-\cos (\theta -s)}{2\sin s}+C\frac{\sin (\theta
+s)-\sin (\theta -s)}{2\sin s}]
=\frac{m(\theta +s)-m(\theta -s)}{2\sin s}. \eqno(S7)
$$

The gradients of $\theta $ with respect to $\Omega $ and $\Delta $ are%
$$
\frac{\partial \theta }{\partial \Omega } =\frac{2\Omega \Delta _{k}\Gamma
}{(\Delta _{k}\delta _{k}-\Omega ^{2})^{2}+(\Delta _{k}\Gamma )^{2}},$$
$$
\frac{\partial \theta }{\partial \Delta } =\frac{\Gamma \Omega ^{2}}{%
(\Delta _{k}\delta _{k}-\Omega ^{2})^{2}+(\Delta _{k}\Gamma )^{2}}. \eqno(S8)
$$

\subsection{Influences of the imperfect chiral couplings and atomic decay}

The left-moving guided photon is inevitably coupled to the atom due to the
manufacturing process. In this case, we label the coupling strength between
the atomic transition and left(right)-moving photon in waveguide $j$ by $%
g_{l,j}(g_{r,j})$, with $g_{l,j}\ll g_{r,j}$. The Hamiltonian turns out to
be
$$
H =-i\sum_{j=1,2}\int dx_{j}[r^{\dagger }(x_{j})\partial
_{x_{j}}r(x_{j})-l^{\dagger }(x_{j})\partial _{x_{j}}l(x_{j})]+(\omega _{2}-\Delta )\sigma ^{33}$$
$$+\omega
_{2}\sigma ^{22}+\sum_{\substack{ j=1,2  \\ h=r,l}}g_{h,j}\int dx_{j}\delta (x_{j}-a)\sigma
^{21}h(x_{j}) +\Omega
\sigma ^{23}+h.c. \eqno(S9)
$$
with $l^{\dagger }(x_{j})$ ($l(x_{j})$) creating (destroying) a left-moving
photon at site $x$ in waveguide $j$. In this case, the dynamics of
left-moving photon should be considered. Therefore, the arbitrary state of
the system becomes%
$$
\left\vert \Psi \right\rangle =\sum_{\substack{ j=1,2  \\ h=r,l}}\int
dx_{j}f_{h,j}(x_{j})h^{\dagger }(x_{j})+l\sigma ^{21}+h\sigma
^{31}\}\left\vert \phi \right\rangle ,\eqno(S10)
$$%
with $f_{h,j}$, $l$ and $h$ corresponding probability amplitudes. By
performing similar calculations, we obtain the transfer matrix as%
$$
\left[
\begin{array}{c}
f_{r,1}(a^{+}) \\
f_{r,2}(a^{+}) \\
f_{l,1}(a^{+}) \\
f_{l,2}(a^{+})%
\end{array}%
\right] =\frac{1}{A}\left[
\begin{array}{cccc}
A+i\Delta _{k}\Gamma _{r,1} & i\Delta _{k}\sqrt{\Gamma _{r,1}\Gamma _{r,2}}
& i\Delta _{k}\sqrt{\Gamma _{r,1}\Gamma _{l,1}} & i\Delta _{k}\sqrt{\Gamma
_{r,1}\Gamma _{l,2}} \\
i\Delta _{k}\sqrt{\Gamma _{r,1}\Gamma _{r,2}} & A+i\Delta _{k}\Gamma _{r,2}
& i\Delta _{k}\sqrt{\Gamma _{r,2}\Gamma _{l,1}} & i\Delta _{k}\sqrt{\Gamma
_{r,2}\Gamma _{l,2}} \\
i\Delta _{k}\sqrt{\Gamma _{r,1}\Gamma _{l,1}} & i\Delta _{k}\sqrt{\Gamma
_{l,1}\Gamma _{r,2}} & A+i\Delta _{k}\Gamma _{l,1} & i\Delta _{k}\sqrt{%
\Gamma _{l,1}\Gamma _{l,2}} \\
i\Delta _{k}\sqrt{\Gamma _{r,1}\Gamma _{l,2}} & i\Delta _{k}\sqrt{\Gamma
_{l,2}\Gamma _{r,2}} & i\Delta _{k}\sqrt{\Gamma _{l,2}\Gamma _{l,1}} &
A+i\Delta _{k}\Gamma _{l,2}%
\end{array}%
\right] \left[
\begin{array}{c}
f_{r,1}(a^{-}) \\
f_{r,2}(a^{-}) \\
f_{l,1}(a^{-}) \\
f_{l,2}(a^{-})%
\end{array}%
\right] , \eqno(S11)
$$%
with $\Gamma _{h,j}=g_{h,j}^{2}$ and $A=\Delta _{k}(\delta _{k}-i\frac{%
\Gamma _{r,1}+\Gamma _{r,2}+\Gamma _{l,1}+\Gamma _{l,2}}{2})-\Omega ^{2}$.

In addition, there is atomic decay to other modes other than waveguides.
This dissipation can be incorporated by introducing the nonhermitian
Hamiltonian $H_{non}=-i\frac{\gamma }{2}\sigma ^{22}$ in the quantum jump
picture, with $\gamma $ the decay rate of level $\left\vert 2\right\rangle $%
. Here we have assumed that levels $\left\vert 1\right\rangle $ and $%
\left\vert 3\right\rangle $ are long-live. Consequently, the label $\delta
_{k}$ in the transfer matrices is replaced by $\delta _{k}-i\frac{\gamma _{2}%
}{2}$. Note the label $\Delta _{k}$ remains unchanged although it contains $%
\delta _{k}$ from the formation below Eq. (S3). This is because,
essentially, $\Delta_{k}=\omega _{3}+\omega _{L}-E$.

In Fig. S1 (a), we numerically simulate the performance of the
single-qubit gate. We assume $\Gamma _{r,1}=\Gamma _{r,2}=\Gamma _{R}$, $%
\Gamma _{l,1}=\Gamma _{l,2}=\Gamma _{L}$. Initially, a photon is injected
into the waveguide $1$. The fidelity is introduced to measure the overlap
between the ideal output state and the output state with the imperfect
chiral coupling and atomic decay.

\begin{figure}[t]
\begin{center}
\includegraphics[width=12cm, height=4cm]{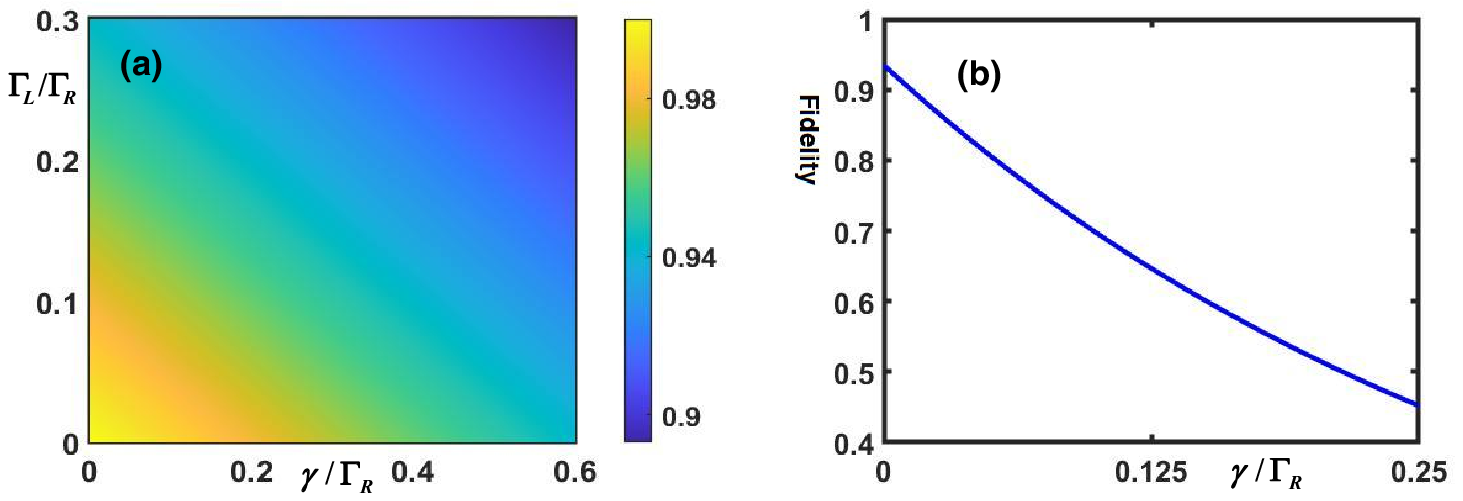}
\end{center}
FIG. S1: Numerical simulation of the fidelity when the imperfect chiral
coupling and atomic decay are considered. (a) denotes the fidelity for the
operation of \emph{Type a}, with the parameters $\Omega=2\Gamma_R$, $%
\Delta_k=\protect\delta_k=\Gamma_R$. (b) denotes the fidelity against the
atomic decay when $\Gamma_L/(\Gamma_R+\Gamma_L)=1/60$ for the operation of
\emph{Type c}. The other parameters of (b) agree with the ones realizing the
C-NOT gate in the main text.
\label{fig1}
\end{figure}

\section{A five-level atom is chirally coupled to three waveguides}

We consider a five-level atom chirally coupled to three waveguides at site $%
a $, \textit{type c} of Fig. 1 in the main text. For conveniently describing
the calculations, in this section, we label the waveguides $m_{1}$,
waveguides $m_{0}^{\prime }$, and waveguides $m_{1}^{\prime }$ by waveguide $%
1$, $2$, and $3$, respectively. The frequencies corresponding to the states $%
\left\vert 1\right\rangle $, $\left\vert 2\right\rangle $, $\left\vert
3\right\rangle $, $\left\vert 4\right\rangle $, and $\left\vert
5\right\rangle $, are denoted by $\omega _{1}$, $\omega _{2}$, $\omega _{3}$%
, $\omega _{4}$ and $\omega _{5}$, respectively. We choose the ground-state
energy for reference, i. e. $\omega _{1}=0$. We assume that the atom is
initialized in the state $\left\vert 1\right\rangle $. The coupling between
the five-level atom and the waveguides act as a controlled-rotating
operation. To illustrate this, we investigate two situations.

\textit{Situation A. }We first inject a right-moving single photon with
energy $E$ into the waveguide $1$. In the single-excitation case, the atomic
level $\left\vert 4\right\rangle $, waveguide $2$ and waveguide $3$\ do not
participate in the dynamic process. Therefore, the system can be considered
as a four-level atom coupled to the waveguide $1$. The time-independent
Hamiltonian has the form%
$$
H =-i\int dx_{1}r^{\dagger }(x_{1})\partial _{x_{1}}r(x_{1})+\omega
_{2}\sigma ^{22}+(\omega _{3}-\omega _{L})\sigma ^{33}+(\omega _{4}-\omega
_{L})\sigma ^{55}+\Omega \sigma
^{23}$$
$$+g\int dx_{1}\delta (x_{1}-a)\sigma ^{12}r^{\dagger }(x_{1})+g^{\prime
}\int dx_{1}\delta (x_{1}-a)\sigma ^{53}r^{\dagger }(x_{1})+h.c., \eqno(S12)
$$
where $\omega _{L}$ is the frequency of the external Laser, $\sigma
^{mn}=\left\vert m\right\rangle \left\langle n\right\vert $ ($m,n=1,2,3,5$)
are the atomic raising, lowering and energy level population operators, and $%
r^{\dagger }(x_{1})$($r(x_{1})$) creates(destroys) a right-moving photon at
site $x$ in waveguide $1$. The arbitrary state is%
$$
\left\vert \Psi \right\rangle =\int dxf(x)r^{\dagger }(x_{1})\left\vert \phi
\right\rangle +\int dxf^{\prime }(x)r^{\dagger }(x_{1})\sigma
^{51}\left\vert \phi \right\rangle +l\sigma ^{21}\left\vert \phi
\right\rangle +h\sigma ^{31}\left\vert \phi \right\rangle \eqno(S13)
$$
where $f(x_{1})$, $f^{\prime }(x_{2})$, $l$ and $h$ are probability
amplitudes. The state $\left\vert \phi \right\rangle $ denotes that the atom
is in the state $\left\vert 1\right\rangle $, and the waveguide $1$ does not
contain any photon. The second term denotes the processing that the atom
absorbs the injected photon and then reemits a photon, meanwhile it makes a
transition $\left\vert 1\right\rangle \rightarrow \left\vert 5\right\rangle $%
. From the scattering \theequation, we obtain%
$$
Ef(x) =-i\partial _{x}f(x)+\sqrt{\Gamma }l\delta (x-a)  \notag$$
$$Ef^{\prime }(x) =-i\partial _{x}f\prime (x)+(\omega _{4}-\omega
_{L})f^{\prime }(x)+\sqrt{\Gamma }h\delta (x-a) $$
$$El =\omega _{2}l+\sqrt{\Gamma }f(a)+\Omega h $$
$$Eh =(\omega _{3}-\omega _{L})h+\sqrt{\Gamma }f^{\prime }(a)+\Omega l
\label{eigenset1} \eqno(S14)
$$%
Here we take $g=g^{\prime }=\sqrt{\Gamma }$. From the second equation in the
equation set (S14), if
the atom is in the state $\left\vert 1\right\rangle $ after scattering, the
energy of the scattered photon is $E$. However, if the atom is in the state $%
\left\vert 5\right\rangle $, the energy of the scattered photon is $%
E-(\omega _{5}-\omega _{L})$. It implies the frequency conversion. By
adjusting the frequency of the Laser, the frequency of the scattered photon
is controlled. The conversion efficiency is represented by $\left\vert
f^{\prime }(x)\right\vert ^{2}$ ($x>a$). By solving the equation set (S14),
we find%
$$
\left[
\begin{array}{c}
f(a^{+}) \\
f^{\prime }(a^{+})%
\end{array}%
\right] =\left[
\begin{array}{cc}
\frac{\delta _{k}\delta _{k}^{\prime }-\Omega ^{2}-i\frac{\Gamma }{2}\delta
_{k}+i\frac{\Gamma }{2}\delta _{k}^{\prime }+\frac{\Gamma ^{2}}{4}}{\delta
_{k}\delta _{k}^{\prime }-\Omega ^{2}-i\frac{\Gamma }{2}\delta _{k}-i\frac{%
\Gamma }{2}\delta _{k}^{\prime }-\frac{\Gamma ^{2}}{4}} & \frac{-i\Gamma
\Omega }{\delta _{k}\delta _{k}^{\prime }-\Omega ^{2}-i\frac{\Gamma }{2}%
\delta _{k}-i\frac{\Gamma }{2}\delta _{k}^{\prime }-\frac{\Gamma ^{2}}{4}}
\\
\frac{-i\Gamma \Omega }{\delta _{k}\delta _{k}^{\prime }-\Omega ^{2}-i\frac{%
\Gamma }{2}\delta _{k}-i\frac{\Gamma }{2}\delta _{k}^{\prime }-\frac{\Gamma
^{2}}{4}} & \frac{\delta _{k}\delta _{k}^{\prime }-\Omega ^{2}+i\frac{\Gamma
}{2}\delta _{k}-i\frac{\Gamma }{2}\delta _{k}^{\prime }+\frac{\Gamma ^{2}}{4}%
}{\delta _{k}\delta _{k}^{\prime }-\Omega ^{2}-i\frac{\Gamma }{2}\delta
_{k}-i\frac{\Gamma }{2}\delta _{k}^{\prime }-\frac{\Gamma ^{2}}{4}}%
\end{array}%
\right] \left[
\begin{array}{c}
f(a^{-}) \\
f^{\prime }(a^{-})%
\end{array}%
\right] , \eqno(S15)
$$%
with $\delta _{k}=\omega _{2}-E$ and $\delta _{k}^{\prime }=\omega
_{3}-\omega _{L}-E$.
Obviously, when
$\delta _{k}\delta _{k}^{\prime }-\Omega ^{2}+\frac{\Gamma ^{2}}{4} =0$
and
$\delta _{k} =\delta _{k}^{\prime }$,
the atom is determinably in the state $\left\vert 5\right\rangle $ after
scattering and the conversion efficiency is unit. For different converted
frequencies, one can adjust the Rabi frequency of the Laser to realize the
efficient frequency conversion. In the main text, we take the resonant case
that $\delta _{k}=\delta _{k}^{\prime }=0$ and $\Omega =\frac{\Gamma }{2}$.

Subsequently, a single photon is injected in to the waveguide 2 or waveguide
3, which drives the atomic transition $\left\vert 4\right\rangle
\leftrightarrow \left\vert 5\right\rangle $ with strength $g^{\prime \prime
} $. This can be considered as a two-level atom coupled to the two
waveguides. The transfer matrix is obtained from the outcome in Eq. (S4) by
shutting the external Laser, i. e.%
$$
T^{\prime \prime }=\left[
\begin{array}{cc}
\frac{\delta _{k}^{\prime \prime }}{\delta _{k}^{\prime \prime }-i\Gamma } &
\frac{i\Gamma }{\delta _{k}^{\prime \prime }-i\Gamma } \\
\frac{i\Gamma }{\delta _{k}^{\prime \prime }-i\Gamma } & \frac{\delta
_{k}^{\prime \prime }}{\delta _{k}^{\prime \prime }-i\Gamma }%
\end{array}%
\right] =\frac{\delta _{k}^{\prime \prime }}{\delta _{k}^{\prime \prime
}-i\Gamma }I+\frac{i\Gamma }{\delta _{k}^{\prime \prime }-i\Gamma }X, \eqno(S16)
$$
with $\delta _{k}^{\prime \prime }$ the detuning between the photon in
waveguide 2 or 3 and the corresponding atomic transition.\ Here we take $%
g^{\prime \prime }=\sqrt{\Gamma }$. In supervised learning, we consider the
resonant case, i. e. $\delta _{k}^{\prime \prime }=0$ and hence the transfer
matrix has the form of $T^{\prime \prime }=-X$.

\textit{Situation B. }We do not inject any photon into the waveguide $1$.
The atom maintains in its state $\left\vert 1\right\rangle $. If a single
photon is injected in to the waveguide 2 or waveguide 3, it is decoupled to
the atom because the atom is not in the state $\left\vert 4\right\rangle $
or $\left\vert 5\right\rangle $.

The influences of the imperfect chiral coupling and dissipation can be
studied in a similar way to the $\Lambda$-type emitter. In Fig. S1
(b), we numerically simulate this influence resulting in the operation of
\textit{Type c }when the target photon is injected in to the waveguide 2. We
have assumed that the levels $\left\vert 1\right\rangle $ and $\left\vert
5\right\rangle $ are long-live, and all the atomic decays from the excited
states are equal, labeled by $\gamma $. It is shown that the performance is
greatly affected by atomic decays because there are three excited states
participating in the dynamics. When the decay is near $0.02\Gamma _{R}$, the
fidelity is near 0.9.